\documentclass[aps,pra,showpacs,twocolumn]{revtex4}

\usepackage{amssymb}
\usepackage{epsfig}
\usepackage{graphicx}
\usepackage{amsmath}
\usepackage{array,color}

\begin{document}

\title{Dynamics and modulation of ring dark soliton in 2D Bose-Einstein \\ condensates with tunable interaction}
\author{Xing-Hua Hu$^1$, Xiao-Fei Zhang$^{1,2}$, Dun Zhao$^{3,4}$, Hong-Gang Luo$^{4,5}$, and W. M. Liu$^1$}
\address{$^1$Beijing National Laboratory for Condensed Matter Physics,
Institute of Physics, Chinese Academy of Sciences, Beijing 100190,
China}
\address{$^2$College of Science, Honghe University, Mengzi 661100, China}
\address{$^3$School of Mathematics and Statistics, Lanzhou University, Lanzhou 730000,
China}
\address{$^4$Center for Interdisciplinary Studies, Lanzhou University, Lanzhou 730000,
China}
\address{$^5$Institute of Theoretical Physics, Chinese Academy of Sciences, P.O. Box 603, Beijing
100080, China}
\date{\today}

\begin{abstract}
We investigate the dynamics and modulation of ring dark soliton in
2D Bose-Einstein condensates with tunable interaction both
analytically and numerically. The analytic solutions of ring dark
soliton are derived by using a new transformation method. For
shallow ring dark soliton, it is stable when the ring is slightly
distorted, while for large deformation of the ring, vortex pairs
appear and they demonstrate novel dynamical behaviors: the vortex
pairs will transform into dark lumplike solitons and revert to ring
dark soliton periodically. Moreover, our results show that the
dynamical evolution of the ring dark soliton can be dramatically
affected by Feshbach resonance, and the lifetime of the ring dark
soliton can be largely extended which offers a useful method for
observing the ring dark soliton in future experiments.
\end{abstract}


\maketitle

\section{INTRODUCTION}

Solitons are fundamental excitations of nonlinear media and have
attracted great interests from diverse contexts of science and
engineering, such as dynamics of waves in shallow water, transport
along DNA and other macromolecules, and fiber optic communications.
The realization of Bose-Einstein condensates (BECs) \cite{bec}
introduces an unparalleled platform for the study of these nonlinear
excitations, where both bright \cite{bright} and dark solitons
\cite{Burger, Denschlag, Dutton, Ginsberg, Anderson} have been
observed. Very recently, quasi-1D dark solitons with long lifetime
up to $2.8$s have been created in the laboratory \cite{Becker}. This
offers an unusual opportunity to study dark solitons and the
relevant theories.

Dark solitons are robust localized defects in  repulsive BECs, which
are characterized by a notch in the condensate density and a phase
jump across the center \cite{Kivshar2}. So far most of the works on
dark soliton are limited to one-dimensional (1D) BECs
(\cite{Gorlitz}, for a review see \cite{Proukakis, Sevrekidis,
Carretero}), where they are both stable and easily controlled in
experiments. However, in two dimensions (2D), most types of  dark
solitons are short lived due to dynamical instability arising from
higher dimensionality \cite{Denschlag}. For example, in ref.
\cite{huang} it is shown that both 2D dark lumplike soliton and dark
stripe soliton will decay into vortex pairs under small transverse
perturbations. Also these 2D defects are strongly affected by
inhomogeneity of the system and suffer from the snaking instability
just as in optical systems (\cite{Tikhonenko, Kivshar2} and
references therein). Therefore their long-time dynamical behaviors
are hard to observe in real experiments.

One candidate for observing the long-time behavior of 2D dark
soliton is ring dark solitons (RDS) which was first introduced in
the context of optics \cite{Kivshar, Baluschev} and then studied in
BECs \cite{Theocharis}. In a 2D nonlinear homogeneous system, it was
first predicted  that the instability band of a dark stripe soliton
can be characterized by a maximum perturbation wavenumber $Q_{max}$
\cite{Kuznetsov}; and if the length of the stripe is smaller than
the inverse of the wavenumber $L<2\pi/Q_{max}$, then the stripe can
be bent into an annulus with the instability being largely
suppressed. This was further confirmed for BECs even in an
inhomogeneous trap, where both oscillatory and stationary ring dark
solitons can exist \cite{Theocharis}. Besides, the symmetry of the
ring soliton determines that it is little affected by the
inhomogeneity of BEC system. Moreover, the study indicated that the
collisions between the RDSs are quasielastic and their shapes will
not be distorted \cite{Nistazakis}. Due to these specific
characteristics of the RDS, it has stimulated great interests on
observing  2D dark solitons in BECs \cite{Carr, xue, Theocharis3,
Dong}. Recently, Yang {\it et al.} suggested a proposal on how to
create the RDSs in experiment \cite{yang}. Nevertheless, the
generation of the RDS and their dynamical behaviors in BECs have not
been observed in real experiment yet. This is because the lifetime
of deep RDS is not long enough for the experimental observation and
the stability analysis for shallow RDS has not been explored
thoroughly.

In this paper, we first develop a general effective analytical
method to derive the solutions of RDS. This is realized by
transforming the Gross-Pitaevskii (GP) equation with trap potential
to a standard nonlinear Schr\"{o}dinger (NLS) equation. Then we
numerically solve the 2D GP equation and obtain a stability diagram
where a stable domain of shallow RDSs can exist. In the unstable
region, we find a novel transformation process between various dark
solitons and vortex pairs. Finally, we show that the lifetime of
RDSs can be largely extended by Feshbach resonance, which is of
particular importance for the experimental observation of RDSs.

\section{The Model}A BEC trapped in an external potential is described by
a macroscopic wave function $\Psi(\textbf{r},t)$ obeying the GP
equation \cite{Dalfovo}, which reads
\begin{equation}
i\hbar\frac{\partial \Psi(\textbf{r},t)}{\partial t} \!=\! [ -
\frac{\hbar^2}{2m} \nabla^2 \!+\! V(\textbf{r},t)  \!+\!
g_0(t)\mid\Psi(\textbf{r},t)\mid^2 ] \Psi(\textbf{r},t), \label{GP}
\nonumber
\end{equation}
where the wave function is normalized by the particle number $N=
\int d\textbf{r}\mid\Psi\mid^2$ and $g_0(t)=4\pi\hbar^2a(t)/m$
represents the strength of interatomic interaction characterized by
the $s$ -wave scattering length $a(t)$, which can be tuned by
Feshbach resonance. The trapping potential is assumed to be
$V(\textbf{r},z)=m(\omega_r^2 r^2+\omega_z^2z^2)/2$, where
$r^2=x^2+y^2$, $m$ is the atom mass, and $\omega_{r,z}$ are the
confinement frequencies in the radial and axial directions
respectively. Further assuming $\Omega\equiv\omega_r/\omega_z\ll1$
such that the motion of atoms in the $z$ direction is essentially
frozen to the ground state ($f(z)$) of the axial harmonic trapping
potential, the system can be regarded as quasi-2D. Then we can
separate the degrees of freedom of the wave function as
$\Psi(\textbf{r},t)=\psi(x, y, t)f(z)$, obtaining the 2D GP
equation:
\begin{equation}
i\hbar\frac{\partial \psi}{\partial t} = - \frac{\hbar^2}{2m}
(\frac{\partial^2}{\partial x^2}+\frac{\partial^2}{\partial
y^2})\psi \!+\! \frac{m}{2}\omega_r^2r^2\psi \!+\! g_0(t)
\eta\mid\psi\mid^2\psi ,\label{reduced} \nonumber
\end{equation}
where
\begin{equation}
\eta\equiv\frac{\int dz\mid f(z)\mid^4}{\int dz\mid f(z )\mid^2}.
\nonumber
\end{equation}

It is convenient to introduce the scales characterizing the trapping
potential: the length, time, and wave function are scaled as
\begin{equation}
x=a_h\tilde{x}, t=\frac{\tilde{t}}{\omega_z},
\psi=\frac{\tilde{\psi}}{a_h\sqrt{4\pi a_0\eta}} \nonumber
\end{equation}
respectively, with $a_h=\sqrt{\hbar/m\omega_z}$ and $a_0$ is a
constant length we choose to measure the time-dependent $s$ -wave
scattering length. Then the 2D GP equation is reduced to a
dimensionless form as
\begin{equation}
i\frac{\partial\psi}{\partial t}=-\frac{1}{2}\nabla^2\psi+
g(t)\mid\psi\mid^2\psi+\frac{1}{2}\Omega^2r^2\psi, \label{2GP}
\end{equation}
where $\nabla^2=\partial^2/\partial x^2+\partial^2/\partial
y^2=\partial^2/\partial r^2+1/r\times\partial/\partial r
+\partial^2/\partial \theta^2$, $\Omega=\omega_r/\omega_z$,
$g(t)=a(t)/a_0$ and the tilde is omitted for simplicity. This is the
basic equation we treat analytically and numerically.

\section{transformation method and analytic Solution} In order to study the dynamics of ring dark soliton, we consider the solution of
Eq. (\ref{2GP}) with circular symmetry, $\psi(r,t)$. In the case of
$g(t) = C$, where $C$ is a nonzero positive constant, the main
difficulty to solve Eq. (\ref{2GP}) is the existence of the last
term, i.e. the trapping potential. Without trap, i.e. $\Omega=0$,
the system is described by the standard NLS equation:
\begin{eqnarray}
i\frac{dQ(R,T)}{dT}&+&\frac{1}{2}(\frac{\partial^2Q(R,T)}{\partial
R^2}+\frac{1}{R}\frac{\partial Q(R,T)}{\partial R})\nonumber\\
&-&C\mid Q(R,T)\mid^2Q(R,T)=0, \label{notrap}
\end{eqnarray}
Under small-amplitude approximation, Eq. (\ref{notrap}) has been
transformed to the famous cylindrical KdV (cKdV) equation by using
the perturbation method \cite{Kivshar, xue}. The cKdV equation is
known to be basic nonlinear equation describing cylindrical and
spherical pulse solitons in plasmas, electric lattices and fluids
(see, e.g., \cite{Infeld} for a review) and its exact solution has
been derived \cite{ckdv}. Therefore the soliton solutions of Eq.
(\ref{notrap}) are gained.

However, there is no effective method to solve Eq. (\ref{2GP})
generally, especially when the last term is time-dependent, $V(t)$.
Now we develop a method which can transform the general form of Eq.
(\ref{2GP}) to Eq. (\ref{notrap}) by using a transformation:
\begin{equation}
\psi(r,t)=Q(R(r,t), T(t))e^{ia(r,t)+c(t)},
\end{equation}
where $R(r, t)$, $T(t)$, $a(r, t)$, and $c(t)$ are assumed to be
real functions and the transformation parameters read
\begin{eqnarray}
R(r,t)&=& \alpha (t)r, \nonumber \\
T(t)&=&\int \alpha^2(t^\prime)dt^\prime +C_0, \nonumber\\
c(t)&=& \frac{1}{2}\ln\frac{\alpha^2(t)}{C}, \nonumber\\
a(r,t)&=& -\frac{1}{2\alpha (t)}\frac{d\alpha (t)}{dt}r^2,
\end{eqnarray}
where $C_0$ is a constant. The condition that such transformations
exist is
\begin{eqnarray}
\frac{1}{\alpha (t)}\frac{d^2\alpha (t)}{dt^2}-\frac{2}{\alpha
(t)^2}(\frac{d\alpha (t)}{dt})^2-\Omega^2=0. \label{condition}
\end{eqnarray}
Under this condition, all solutions of Eq. (\ref{notrap}) can be
recast into the corresponding solutions of Eq. (\ref{2GP}). So we
build a bridge between the extensive RDS study in nonlinear optics
(homogeneous system) and the trapped BEC system. Furthermore, it is
worth while to note that the transformation method can be used to
solve the general equation:
\begin{eqnarray}
&i&\frac{du(r,t)}{dt}+ D(t)(\frac{\partial^2u(r,t)}{\partial
r^2}+\frac{1}{r}\frac{\partial u(r,t)}{\partial r})\nonumber\\
&+& g(t)\mid u(r,t)\mid^2u(r,t)+V(t)r^2u(r,t)=0, \label{general}
\end{eqnarray}
when $g(t)$ is proportional to $D(t)$. Eq. (\ref{general})
completely describes the dynamics and modulation of both electric
field in optical systems and macroscopic order parameter in atomic
BECs in quasi-2D with circular symmetry.

In order to get the transformation condition explicitly, we
substitute $y(t)=d\alpha/(\alpha dt)$ into Eq. (\ref{condition}) and
obtain
\begin{eqnarray}
\frac{dy(t)}{dt}=y^2+\Omega^2.
\end{eqnarray}
This is a standard Riccati equation. We can solve it not only for
$\Omega=constant$, but also for various type of $\Omega^2(t)$, such
as $a+bsin(\lambda t)$, $a+bcosh(t)$, $ae^{\lambda t}$ ($a$, $b$,
$\lambda$ are arbitrary constants) and so on. Thus we can study the
dynamics of system with time-dependent external trap. Particularly,
when $\Omega$ is independent on $t$, the solution of Eq.
(\ref{condition}) is
\begin{eqnarray}
\alpha(t)=C_1\sec(\Omega t+C_2), \label{solution_c}
\end{eqnarray}
where $C_1$ and $C_2$ are the integral constants.

Since under small-amplitude approximation, the solutions of Eq.
(\ref{notrap}) have been given, then combining the solutions of Eqs.
(\ref{notrap}) and (\ref{solution_c}), we get the corresponding
solutions of RDS in the BEC in the external trap potential. They are
the exact solutions of system when the depth of RDS is infinitely
small, which help us to understand the dynamics of RDS. However,
when the RDSs get deeper, the small-amplitude approximation is
invalid and we have to appeal to the numerical simulation. In the
following sections, we study the dynamics and stability of RDS
numerically.

\section{Stability of shallow ring dark soliton} It has been known that starting from
the initial configuration with strict circular symmetry, the RDS
will oscillate up to a certain time till instabilities develop:
shallow RDS slowly decays into radiation and for deep one, snaking
sets in, leading to formation of vortex-antivortex pairs arranged in
a robust ringshaped array (vortex cluster), because of transverse
perturbations \cite{Theocharis}. But its stability to the
perturbation in the radial direction is not analyzed. Here we study
the stability of shallow RDS against the small distortion of the
ringshape numerically, which is ineluctable in the process of the
practical experiment.

\begin{figure}[tbp]
\centering
\renewcommand{\figurename}{FIG. }
\includegraphics[height=5cm,width=8cm]{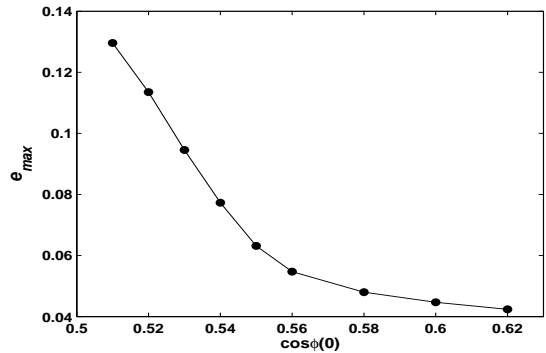}
\caption{ A part of stability diagram of shallow ring dark soliton
with $R_0=28.9$ and the trap frequency $\Omega=0.028$. The maximum
eccentricity $e_{max}$ of the ring in the initial configuration
decreases as the initial depth $\cos\phi(0)$ of the ring dark
soliton increases.}
\end{figure}

We study the stability of RDSs by solving Eq. (\ref{2GP})
numerically with the parameter: $g(t)=1$, $\Omega=0.028$ and the
initial radius of RDS $R_0=28.9$. Because of large initial radius, a
reasonable and good approximation is
\begin{eqnarray}
\psi(x, y, 0)&=&(1-\Omega^2r^2/4)\nonumber\\
&\times&\left[\cos\phi(0)\tanh Z(r_1)+i\sin\phi(0) \right],
\label{IC}
\end{eqnarray}
where $r=\sqrt{x^2+y^2}$, $Z(r_1)=(r_1-R_0)\cos\phi(0)$,
$r_1=\sqrt{(1-e_c^2)x^2+y^2}$, $e_c$ is the eccentricity of the ring
and $\cos\phi(0)$ is proportional to the depth of the input soliton.
When $e_c\neq0$, $R_0$ represents the length of semiminor axis of
the elliptical configuration. We take Eq. (\ref{IC}) as the initial
configuration, the validity of which has been explained in detail
and checked in the previous investigations \cite{Theocharis,
Kivshar}.

\begin{figure}[tbp]
\centering
\renewcommand{\figurename}{FIG. }
\includegraphics{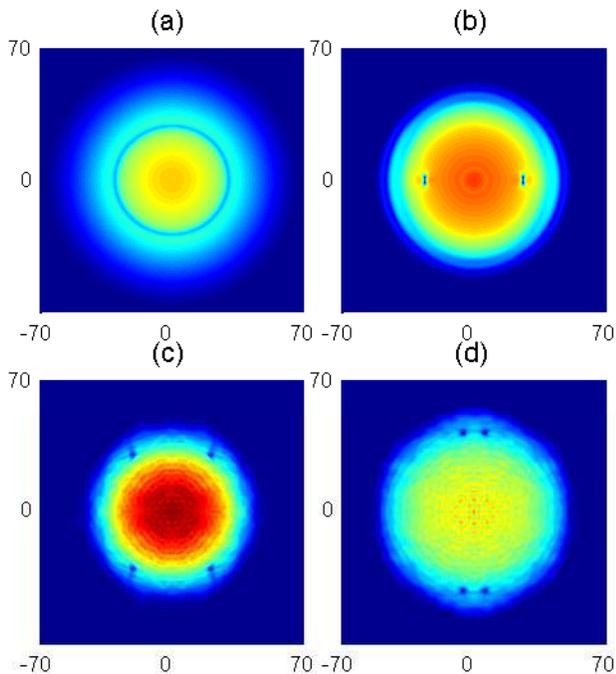}
\caption{(Color online) Evolution of the ring dark soliton with
initial depth $\cos\phi(0)=0.6$, eccentricity $e_c=0.4$, interaction
strength $g(t)=1$ and the trap frequency $\Omega=0.028$. (a) The
initial profile shown by color-scale density plots, where ring dark
soliton corresponds a light-color plot ellipse with the length of
semiminor axis $R_0=28.9$. (b)-(d) correspond to $t=80$, 400, 540
respectively. The ring dark soliton with distortion firstly breaks
into two lump solitons, and then two pairs of vortex pairs. See text
for details. }
\end{figure}

We propagate the 2D time-dependent GP equation (Eq. (\ref{2GP}))
using two distinct techniques: alternating-direction implicit method
\cite{Press, Kasamatsu} and time-splitting Fourier spectral method
\cite{bao}. The results from these two methods are crosschecked and
When $e_c=0$, the properties of results are in agreement with those
in \cite{Theocharis} very well.

To translate the results into units relevant to the experiment
\cite{Burger, Denschlag}, we assume a $^{87}Rb$ ($a_0=5.7$ nm)
condensate of radius $30$ $\mu m$, containing $20 000$ atoms in a
disk-shaped trap with $\omega_r=2\pi\times 18$ Hz and
$\omega_z=2\pi\times 628$ Hz. In this case, the RDS considered above
has the radius $R_0=12.38$ $\mu m$, and the unit of time is $0.25$
ms.

Our numerical results show that the shallow RDS (refers to
$\cos\phi(0)<0.67$, where the soliton without distortion will not
suffer from the snaking instability) is stable against the small
distortion of the ringshape. There is a maximum eccentricity
$e_{max}$ for a given initial depth. When $e_c<e_{max}$, the RDSs
are stable and oscillate reserving their shape with the same period
as the unperturbed RDSs until decaying out. The $e_{max}$ becomes
smaller with the initial depth increasing (see Fig. 1). This is
because $2\pi/Q_{max}$ decreases as the RDS becomes deeper
\cite{Kivshar}, thus the shallower RDS can support larger
distortion.

When $e_c$ exceeds $e_{max}$, snaking sets in and the dark soliton
breaks into two vortex pairs, presenting a striking contrast to the
multiples of four pairs reported previously. And the evolution of
vortex pair is very different from the case of the deep RDSs without
distortion.

\begin{figure}[tbp]
\centering
\renewcommand{\figurename}{FIG. }
\includegraphics{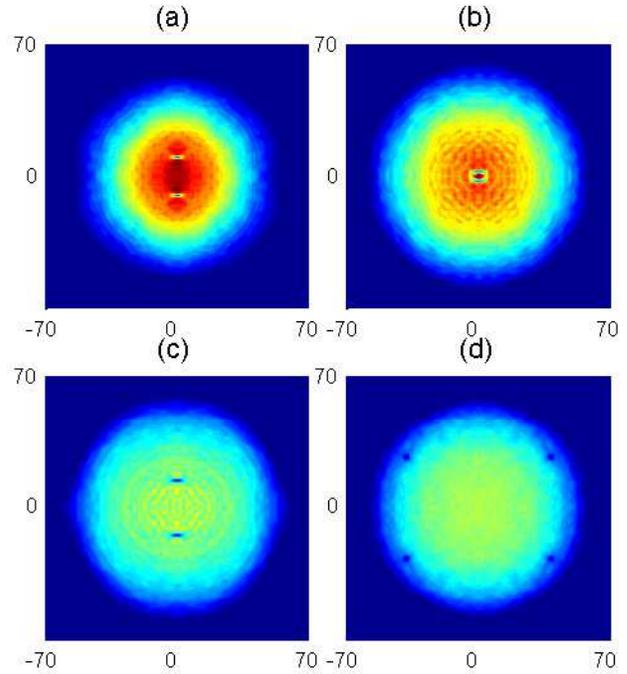}
\caption{(Color online) Evolution of vortex pairs following the Fig.
2 for (a) $t=640$, (b) $t=660$, (c) $t=680$ and (d) $t=1000$. The
vortex pairs collide and merge into dark lumplike solitons, which
move, then form a ring dark soliton for a short time and revive.
Finally the lumplike solitons again break into vortex pairs. See
text for details.}
\end{figure}

To illustrate the generic scenarios, we take the typical case with
$\cos\phi(0)=0.6$ and $e_c=0.4$. It initially shrinks and when
reaching the minimum radius in the short-axis direction, it starts
snaking and forming two dark lumplike solitons in the horizontal
direction; they move in the opposite direction and then break into
two vortex pairs (see Fig. 2(b)). The vortex pairs arrange
themselves in a ring configuration which performs slow radial
oscillations. Simultaneously the vortices and antivortices move
along the ring (see Fig. 2(c, d)).  The result of motion is their
collision in pairs in the vertical direction, followed by vortices
and antivortices's mergence and totally forming a pair of lumplike
solitons (see Fig. 3(a)), which process has been confirmed by phase
distribution of the system. The lumplike solitons move towards to
each other, i.e. to the center of condensate, in the vertical
direction. When they reach the minimum radius, a ring dark soliton
forms (see Fig. 3(b)). After a short time the system returns to the
state of two dark lumplike solitons. The lumplike solitons leave
each other (see Fig. 3(c)) and in the path, each of lumplike
solitons breaks into a vortex pair again. But the configuration is
different to the initial one: the vortex (antivortex) is substituted
by antivortex (vortex), just like the vortex and the antivortex
passing each other directly following the motion before the
mergence. Then the vortices and the antivortices keep on moving (see
Fig. 3(d)), then collide and merge in the horizontal direction. The
dynamical process repeats itself.

This dynamical behavior of the vortex pairs is novel and part of the
process is similar to the behavior of lump soliton reported in
\cite{huang}. The certain centrical collision of vortex pair
provides a potential tool to study the collision dynamics of
vortices.

\section{controlling of deep ring dark solitons}
The deep ring dark solitons (refer to $\cos\phi(0)>0.67$) suffer
from the snaking instability and can only survive for a short time
(typically $10$ ms in the numerical simulation) before changing into
other soliton type. Longer lifetime is necessary for the practical
observation of complex soliton physics such as oscillations or
collisions \cite{Becker}. Our results show that the Feshbach
resonance management technique can largely extend the lifetime of
the ring dark solitons, which makes it possible to study the
long-time behavior of 2D dark soliton in experiment.

Feshbach resonance \cite{Inouye} is a quite effective mechanism that
can be used to manipulate the interatomic interaction (i.e. the
magnitude and sign of the scattering length), which has been used in
many important experimental investigations, such as the formation of
bright solitons \cite{bright}. Feshbach resonance management refers
to the time-periodic changes of magnitude and/or sign of scattering
length by Feshbach resonance \cite{Kevrekidis}, and has been widely
used to modulate and stabilize bright solitons \cite{Ueda, liu},
while its effect on the dynamics and stability of dark soliton is
not very definite \cite{Malomed}. Taking the ring dark soliton as an
example, we demonstrate that the Feshbach resonance can dramatically
affect the dynamics of dark solitons.

\begin{figure}[tbp]
\centering
\renewcommand{\figurename}{FIG. }
\includegraphics{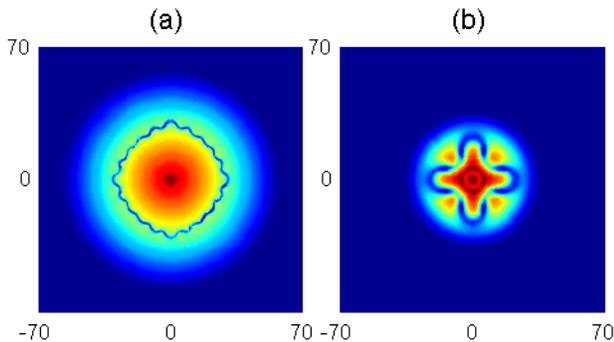}
\caption{(Color online) The snapshots of condensate with ring dark
soliton under the Feshbach resonance management. The parameter are
given as follow: (a) $e_c=0$, $\cos\phi(0)=1$, $g(t)=2-\sin(2\Omega
t)$ and $t=60$ ($\approx15$ ms); (b) $e_c=0$, $\cos\phi(0)=1$,
$g(t)=e^{-\Omega t}$ and $t=100$ ($\approx 25$ ms). Other parameters
are the same as Fig. 2. In the former case the ring dark soliton
will break into $20$ vortex pairs, while only $4$ vortex pairs in
the later one. The total atom number is the same in these two plot,
so for clearness, we use different color scale.}
\end{figure}

To show the effect of Feshbach resonance management, Eq. (\ref{2GP})
is integrated numerically with eccentricity $e_c=0$ and a time
dependent nonlinear coefficient $g(t)$; other parameters and the
initial condition are the same as in Sec.
\uppercase\expandafter{\romannumeral4}. We consider the cases of
$g(t)\sim ae^{\pm\omega t}$ and $g(t)\sim a\pm \sin(\omega t)$,
where $a$ and $\omega$ are arbitrary constant parameters. Our
results show that for the deep ring dark solitons, the Feshbach
resonance management remarkably changes the evolution and the
instability of solitons, indicated by the number of vortex pairs
which arise due to the snaking instability. Two typical examples are
shown in Fig. 4: in the former one, the black ring dark solitons
(refer to $\cos\phi(0)=1$) break into $20$ vortex pairs, while the
latter one only has $4$ pairs. These phenomena are very different
from those with constant scattering length which have $16$ vortex
pairs. This makes it possible to study the dynamics of snaking
instability in detail. What is more interesting and more important
are the cases in which the lifetime of the RDS  can be extended,
such as $g(t)=1-\sin(\Omega t)$, as discussed below.

When the system subjects to the modulation of $g(t)=1-\sin(\Omega
t)$, the RDS can exist for longer time before breakup. For example,
when $\cos\phi(0)=0.76$, the lifetime can be extended up to $45$ ms
from  $10$ ms, in which process, we can observe one complete cycle
of oscillation of RDS. While if we just reduce the scattering length
but keeping it constant, the lifetime of RDS will not change much.
For example, if $g(t)=0.2$, the RDS will become snaking at about
$15$ ms. So the lifetime extension effect is due to the Feshbach
resonance management. From Eq. (\ref{solution_c}), we can see that
$\Omega$ is one intrinsic frequency of system. So it is reasonable
to deduce that this effect is a resonance phenomenon. Furthermore,
this effect is even valid for shallow RDS suffering from the
instability due to the distortion of ringshape. It has been check
that under Feshbach resonance management $g(t)=1-\sin(\Omega t)$,
the RDS with $\cos\phi(0)=0.6$ and $e_c=0.4$, has a lifetime of $50$
ms, which is much larger than the original $10$ ms. Thus under large
perturbation the RDS can exist long enough to be observed, which
make the experimental study of RDS easier.

\section{CONCLUSION}
In conclusion, we have introduced a new transformation method to
derive the solution of ring dark soliton, which provides a powerful
analytic tool to study the 2D BECs and nonlinear optical systems
with circular symmetry. Then we check the stability of shallow ring
dark soliton, and find a novel dynamical behavior mode of solitons.
Furthermore, we study the effect of Feshbach resonance management on
the evolution and the stability of ring dark soliton, also discover
a method to extend the lifetime of ring dark soliton largely. We
show that the ring dark solition is a potential candidate for
observing the long-time behavior of 2D dark soliton.

\section{Acknowledgments}
We acknowledge the heuristic discussion with Weizhu Bao. This work
was supported by NSFC under grants Nos. 60525417, 10740420252,
10874235, the NKBRSFC under grants Nos. 2005CB724508 and
2006CB921400, and the Program for NCET.

\end{document}